\documentclass[11pt,a4paper]{article}
\usepackage{amsmath,amssymb,hyperref,epsfig,multicol}
\newcommand{\alg}[1]{\mathfrak{#1}}
\newcommand{\smat}{\mathcal{S}}
\newcommand{\rmat}{\mathcal{R}}
\newcommand{\beq}{\begin{equation}}
\newcommand{\eeq}{\end{equation}}
\newcommand{\bea}{\begin{eqnarray}}
\newcommand{\eea}{\end{eqnarray}}
\newcommand{\kchi}{|\chi\rangle}
\newcommand{\bchi}{\langle\chi|}
\newcommand{\kphi}{|\phi\rangle}
\newcommand{\bphi}{\langle\phi|}
\newcommand{\kpsi}[1]{|\psi^{#1}\rangle}
\newcommand{\bpsi}[1]{\langle\psi^{#1}|}
\usepackage{epsfig,multicol}
\begin{document}
\begin{titlepage}
\begin{flushright} 
HU-EP-06/22\\
\end{flushright}
\mbox{ }  \hfill hep-th/0608038
\vspace{5ex}
\Large
\begin {center}     
{\bf On the Hopf algebra structure of 
the AdS/CFT S-matrix}
\end {center}
\large
\vspace{1ex}
\begin{center}
Jan Plefka, Fabian Spill and Alessandro Torrielli
\end{center}
\vspace{1ex}
\begin{center}
Humboldt--Universit\"at zu Berlin, Institut f\"ur Physik\\
Newtonstra\ss e 15, D-12489 Berlin, Germany\\[2mm] 
 \vspace{1ex}
\texttt{plefka,spill,torriell@physik.hu-berlin.de}

\end{center}
\vspace{4ex}
\rm
\begin{center}
{\bf Abstract}
\end{center} 
\normalsize 
We formulate the Hopf algebra underlying the $\alg{su}(2|2)$ worldsheet S-matrix of
the $AdS_5\times S^5$ string in the AdS/CFT correspondence. For this we extend the 
previous construction in the $\alg{su}(1|2)$ subsector due to Janik 
to the full algebra by specifying the 
action of the coproduct and the antipode on the remaining generators. 
The nontriviality of 
the coproduct is determined by length-changing effects and results in an 
unusual central braiding. 
As an application we explicitly determine the antiparticle representation by means
of the established antipode.     

\vfill
\end{titlepage} 

\section{Introduction}

Integrable structures play a very important role in the most recent 
developments concerning the AdS/CFT correspondence. After the discovery 
of one-loop integrability of the planar dilatation operator 
in the scalar sector
of four-dimensional
${\cal{N}}=4$ SYM theory \cite{MZ}, important progress has been made in uncovering 
integrable structures in both the gauge and the string theory side of the correspondence. 
On the gauge theory side integrability could be shown to persist to 
higher orders in perturbation theory as well as to the full set of 
field excitations \cite{Gauge}. 
Analogously, integrability has been established at the classical level of string 
theory \cite{BK} and first steps towards quantum integrability have been 
undertaken \cite{quant}. As a result of 
this work, the S-matrix of the problem has been fully determined up to an overall
scalar factor, which seems to be not constrained completely by symmetry 
arguments and whose form should interpolate between the known expressions
obtained perturbatively on both sides of the correspondence. 

The S-matrix of AdS/CFT was shown to possess a centrally extended 
$\alg{su}(2|2) \oplus \alg{su}(2|2)$ symmetry by Beisert \cite{Beisert}, where
the central extensions are related to length-changing effects of an underlying
dynamic spin chain. 
It is 
completely fixed up to the abovementioned scalar factor, and can be written in a manifestly
$\alg{su}(1|2)$ symmetric form, in terms of a combination of projectors      
into irreducible representations of the tensor product 
$\alg{su}(1|2) \otimes \alg{su}(1|2)$. On the other hand, the undetermined phase factor contains an exponential of quadratic terms in the conserved charges of the theory, 
with coefficient functions depending on the `t Hooft coupling \cite{BKl}.  
First orders in the expansion of these coefficients can be determined 
comparing with perturbation theory \cite{dress}. 
Remarkable progress in constraining the scalar factor was made by Janik, who advocated an underlying 
Hopf algebra structure from which he derived an equation analog to the 
crossing symmetry condition of relativistic S-matrices \cite{Janik}. In order to derive
this equation he made
use of the antipode acting on $\alg{su}(1|2)$ generators, where one can 
avoid length-changing effects. The resulting conditions, together with 
unitarity, put strong constraints on the form of the scalar factor. This equation was checked against the known perturbative expansions and agreement was found \cite{AF} (see also \cite{ultBei}).

Unfortunately, these equations do not constrain the overall factor completely, and new insights, in particular on understanding its analytic structure, are needed \cite{Jatalk}.  
In this note we take a step further and establish the Hopf algebra 
structure underlying the {\it full} $\alg{su}(2|2)$ symmetry algebra. This can be 
done by simply analyzing the condition of invariance of the $\alg{su}(1|2)$
S-matrix under the remaining dynamic $\alg{su}(2|2)$ generators. The 
effect of length-changing operators produces momentum dependent phase factors
which can be mathematically represented through a nontrivial bialgebra coproduct, which
we determine. We derive the 
antipode and verify the whole set of Hopf algebra axioms. 
By using the derived antipode we are able to fully 
determine the antiparticle representation directly from the Hopf algebra
structure.

The most important step beyond this would be to determine the universal 
R-matrix of the problem. The requirement of the quasi- 
triangularity condition would lead to an expression of the universal R-matrix
which would solve the crossing relation directly at the algebraic level,
without making use of representation-dependent coefficients. This could be a 
very powerful and clean way of fixing the structure of the overall factor,
and might teach us a lot about the origin of this symmetry from the string theory 
perspective.

\section{Deriving the Coproduct}

The relevant centrally extended $\alg{su}(2|2)$ commutation relations read \cite{Beisert} 
\begin{eqnarray}
\label{algebra}
[\alg{R}^a_b, \alg{J}^c] &=& \delta^c_b \alg{J}^a - \frac{1}{2} \delta^a_b \alg{J}^c,\nonumber\\
\relax
[ {\alg{L}}^\alpha_\beta, \alg{J}^\gamma ] &=& \delta^\gamma_\beta \alg{J}^\alpha - \frac{1}{2} \delta^\alpha_\beta 
\alg{J}^\gamma,\nonumber\\
\{\alg{Q}^\alpha_a,\alg{S}^b_\beta\} &=& \delta^b_a \alg{L}^\alpha_\beta + \delta^\alpha_\beta \alg{R}^b_a +  
\delta^b_a  \delta^\alpha_\beta \alg{C} \nonumber\\
\{\alg{Q}^\alpha_a,\alg{Q}^\beta_b\} &=& \epsilon^{\alpha\beta}\epsilon_{ab}\alg{P}, \nonumber\\
\{\alg{S}^a_\alpha,\alg{S}_\beta^b\} &=& \epsilon_{\alpha\beta}\epsilon^{ab}\alg{K}\, .
\end{eqnarray}
Here $\alg{J}$ represents any generator with the appropriate index, and the $\alg{R}$'s and $\alg{L}$'s close on two copies of $\mathfrak{su}(2)$ 
respectively among themselves. The central charges are $\alg{C}, \alg{P}$ and $\alg{K}$.

Working in the $\alg{su} (1|2)$ language, the S-matrix can be determined as a 
combination of projectors into the irreducible representations of the tensor
product $\alg{su}(1|2) \otimes \alg{su}(1|2)$, weighted with 
representation-dependent coefficients. 
Invariance 
of the S-matrix under all $\alg{su} (1|2)$ generators amounts in fact 
to impose a trivial
coproduct condition. More specifically, let us indicate collectively
with ${\alg{T}}$ any of the $\alg{su} (1|2)$ generators ${\alg{R}}^1{}_1$,
${\alg{R}}^2{}_2 = - {\alg{R}}^1{}_1$, 
${\alg{L}}^\alpha{}_\beta$, ${\alg{Q}}^1{}_1$,
${\alg{Q}}^2{}_1$, ${\alg{S}}^1{}_1$ and ${\alg{S}}^1{}_2$, together with the 
central charge ${\alg{C}}$. The ambient space we set is the universal 
enveloping algebra $U(\alg{su}(2|2))$ containing generators of 
the Lie algebra and all their products. Together with the (undeformed) 
defining commutation relations, and the standard notion of unit, which define the multiplication structure, we will endow this space with a coalgebra structure by specifying the coproduct and the counit, making it a bialgebra. Finally, the antipode will determine the Hopf algebra structure on $U(\alg{su}(2|2))$.

Invariance of the S-matrix under $\alg{T}$ amounts to the following
condition \cite{Beisert}

\begin{equation}
\label{1}
[{\alg{T}}_1 + {\alg{T}}_2, \smat_{12}] = 0.
\end{equation}
This can be rewritten in terms of an R-matrix 

\begin{equation}
\label{1a}
\smat = \sigma \circ \rmat,
\end{equation}
where $\sigma (A \otimes B) = {(-1)}^{deg(A)deg(B)} B \otimes A$, as a coproduct relation

\begin{equation}
\label{2}
\Delta^{op}({\alg{T}}) \, \rmat = \rmat \, \Delta({\alg{T}}),
\end{equation}
with the coproduct $\Delta$ defined as

\begin{equation}
\label{3}
\Delta({\alg{T}}) = ({\alg{T}} \otimes id + id \otimes {\alg{T}})
\end{equation}
and $\Delta^{op} = \sigma \circ \Delta$.
One recalls that $\smat : V_1 \otimes V_2 \rightarrow V_2 \otimes V_1$, $V_i$ 
being modules for $\alg{su}(2|2)$ representations. 
In order to get eq.~(\ref{1}) from eqs.~(\ref{2}) and (\ref{3}) one has simply 
to project  the two factors 
of the tensor product in the abstract algebra onto some definite representations 
$1$ and $2$. We recall that a bialgebra 
with the property that the opposite coproduct $\Delta^{op}$ is equal to the coproduct
$\Delta$ is referred to as ``co-commutative''. 

An R-matrix of the form 

\begin{equation}
\label{4}
\rmat_{12} = \sum_i S_i P_i, 
\end{equation}
where $P_i$ are 
the projectors into the irreducible representations of 
$\alg{su} (1|2) \otimes \alg{su} (1|2)$, and $S_i$ are arbitrary coefficients,
solves eq.~(\ref{1}). There are three such projectors, whose expression
in terms of the quadratic $\alg{su} (1|2)$ Casimir can be found in 
\cite{Janik}. The coefficients $S_i$ are then fixed by requiring invariance 
under 
the remaining generators of $\alg{su} (2|2)$ which are not in the 
$\alg{su} (1|2)$ subalgebra\footnote{In a quite different context, this is however similar to the standard Jimbo-equation procedure for determining the R-matrix for quantum affine algebras, see for instance\cite{Delius}.}, namely ${\alg{R}}^2{}_1$, ${\alg{R}}^1{}_2$,
${\alg{Q}}^1{}_2$,
${\alg{Q}}^2{}_2$, ${\alg{S}}^2{}_1$ and ${\alg{S}}^2{}_2$, together with the 
central charges ${\alg{P}}$ and ${\alg{K}}$. These generators 
are called ``dynamic'' and we collectively denote them by $\alg{D}$.
All these generators change the 
length of the spin chain when acting on all magnons. In order to close 
their action on states of a same chain, 
one can use the basic relation (2.13) 
of \cite{Beisert} to   
move all length-changing operators to the right of all excitations, 
and exploit the limit of having an
infinite chain. This produces the appearance of braiding factors. For example, 
invariance under ${\alg{Q}}^1{}_2$ leads to eq.~(34) of \cite{Janik},

\begin{equation}
\label{5}
(e^{-i p_1} \, {[\tilde{{\alg{Q}}}^1{}_2]}_2 \otimes id_1 + 
id_2 \otimes {[\tilde{{\alg{Q}}}^1{}_2]}_1) \smat = \smat 
(e^{-i p_2} \, {[\tilde{{\alg{Q}}}^1{}_2]}_1 \otimes id_2 + 
id_1 \otimes {[\tilde{{\alg{Q}}}^1{}_2]}_2),
\end{equation}
where the subscript indicates the representations, and
the tilded version of the operator means the same action as the untilded 
one but disregarding length-changing effects, which are taken into account 
by the 
braiding factors. In this case one defines (cfr. \cite{Janik})

\begin{equation}
\label{6}
\tilde{{\alg{Q}}}^1{}_2 = a  \kpsi{1}\bchi - b  \kphi\bpsi{2}. 
\end{equation}
Now note that one can rewrite eq.~(\ref{5}) in terms of a deformed coproduct

\begin{equation}
\label{7}
\Delta^{op}(\tilde{{\alg{Q}}}^1{}_2) \, \rmat = \rmat \, 
\Delta(\tilde{{\alg{Q}}}^1{}_2),
\end{equation}
with the coproduct $\Delta$ defined as

\begin{equation}
\label{8}
\Delta(\tilde{{\alg{Q}}}^1{}_2) = (\tilde{{\alg{Q}}}^1{}_2 
\otimes e^{- i p} + id \otimes \tilde{{\alg{Q}}}^1{}_2).
\end{equation}
We have lifted the coproduct relation to a representation-independent level,
where we understand now $e^{- i p}$ as a central 
element of the (universal enveloping algebra of the) $\alg{su} (2|2)$
algebra. One could make use of the natural choice 

\begin{equation}
\label{nat1}
e^{- i p} = 1 + \frac{1}{\alpha} {\alg{P}},
\end{equation}
imposed by physical requirement, 
where $\alpha$ is a parameter related to the coupling constant \cite{Beisert}, and another similar 
relation for 

\begin{equation}
\label{nat2}
e^{i p} = 1 + \frac{1}{\beta} {\alg{K}}.
\end{equation}
 
One can immediately realize that the coproduct (\ref{8}) is not co-commutative. The existence of an element $\rmat$ of the tensor algebra 
$U(\alg{su} (2|2)) \otimes U(\alg{su} (2|2))$ such that (\ref{7}) holds
makes the bialgebra ``quasi-cocommutative''. The presence of $e^{- i p}$ is connected to the fact that the dynamic
generator ${\alg{Q}}^1{}_2$ {\it adds} one ${\cal{Z}}$ field to the chain.
Let us spell out the coproducts one obtains in an analog way
for all the other dynamic generators :

\begin{eqnarray}
\label{10}
&&\Delta(\tilde{{\alg{Q}}}^2{}_2) = (\tilde{{\alg{Q}}}^2{}_2 
\otimes e^{- i p} + id \otimes \tilde{{\alg{Q}}}^2{}_2),\nonumber\\
&&\Delta(\tilde{{\alg{S}}}^2{}_1) = (\tilde{{\alg{S}}}^2{}_1 
\otimes e^{i p} + id \otimes \tilde{{\alg{S}}}^2{}_1),\nonumber\\ 
&&\Delta(\tilde{{\alg{S}}}^2{}_2) = (\tilde{{\alg{S}}}^2{}_2 
\otimes e^{i p} + id \otimes \tilde{{\alg{S}}}^2{}_2),\nonumber\\
&&\Delta(\tilde{{\alg{R}}}^1{}_2) = (\tilde{{\alg{R}}}^1{}_2 
\otimes e^{- i p} + id \otimes \tilde{{\alg{R}}}^1{}_2),\nonumber\\
&&\Delta(\tilde{{\alg{R}}}^2{}_1) = (\tilde{{\alg{R}}}^2{}_1 
\otimes e^{i p} + id \otimes \tilde{{\alg{R}}}^2{}_1).
\end{eqnarray}
One can notice the conjugate braiding $e^{i p}$ for the generators
{\it subtracting} one ${\cal{Z}}$ field from the chain. 

The central charges ${\alg{P}}$ and ${\alg{K}}$ also add length-changing 
operators to all states. Therefore, their coproduct should also be 
deformed in the following fashion. 

\begin{eqnarray}
\label{12}
&&\Delta(\tilde{{\alg{P}}}) = (\tilde{{\alg{P}}} 
\otimes e^{- i p} + id \otimes \tilde{{\alg{P}}}),\nonumber\\
&&\Delta(\tilde{{\alg{K}}}) = (\tilde{{\alg{K}}}
\otimes e^{i p} + id \otimes \tilde{{\alg{K}}}).
\end{eqnarray}
Making use of the relations (\ref{nat1}) and (\ref{nat2}) 
one realizes that 
\begin{equation}
\Delta^{op}(\tilde{\alg{P}}) = \Delta(\tilde{\alg{P}})
\qquad
\Delta^{op}(\tilde{\alg{K}}) = \Delta(\tilde{\alg{K}})
\end{equation}
Together with their centrality, this makes the coproduct relation 
with the R-matrix automatically satisfied 
$
\Delta^{op}(\tilde{\alg{P}})\, \rmat =  \Delta(\tilde{\alg{P}})\,  \rmat =\rmat\, \Delta(\tilde{\alg{P}})
$
and identically for $\tilde{\alg{K}}$. Conversely, the requirement of co-commutativity for the central elements at the abstract algebraic level automatically enforces the quartic constraint \cite{Beisert,Janik}.

\section{The deformed Hopf Algebra structure}

We have seen in the previous section that, in order to implement the length-changing effect at the algebraic level, one can deform the universal enveloping algebra of the symmetry algebra. We do it in such a way that the ordinary commutation relations remain unchanged, therefore only the coalgebra structure gets modified. This produces a well defined bialgebra which we equip with an antipode, making it a Hopf algebra. We would like now the check the Hopf algebra axioms for our construction. The reader is referred to \cite{books} for standard textbooks on Hopf algebras. 

\subsection{The Coproduct}

Since one has no length-changing effects on the $\alg{su}(1|2)$ sector, one defines the coproduct to be the trivial one:

\beq
\Delta(J) = 1\otimes J + J\otimes 1 \qquad \forall J\in \alg{su}(1|2).
\eeq

For $D\in \alg{su}(2|2)/\alg{su}(1|2)$ we need to introduce a braiding $B(D)$  to account for those length-changing effects (tildes on generators are implicitly understood): 

\beq
\Delta(D) = 1\otimes D + D\otimes B(D). 
\eeq

In the following we check that this indeed gives a consistent Hopf algebra structure, provided that we also modify the antipode $S$, and that there are consistency relations between the different braiding factors $B(D)$.
Looking at the commutation relations (\ref{algebra}) of $\alg{su}(2|2)$, and naming the sets ${\cal{J}} = \alg{su}(1|2)$ and ${\cal{D}} = \alg{su}(2|2)/ \alg{su}(1|2)$ we see that
\begin{eqnarray}
&&[{\cal{J}},{\cal{J}}] \subseteq {\cal{J}}, \\
&&[{\cal{J}},{\cal{D}}] \subseteq {\cal{D}}, \\
&&[{\cal{D}},{\cal{D}}] \subseteq {\cal{J}}.
\end{eqnarray}
Now let $J\in \mathfrak{su}(1|2)$,  $D\in \mathfrak{su}(2|2)/\mathfrak{su}(1|2)$, then $W:= [J,D] \in \mathfrak{su}(2|2)/\mathfrak{su}(1|2)$.
The coproduct is required to be a homomorphism, that is, it has to respect the commutation relations\footnote{$[A,B]$ denotes the usual supercommutator: $[A,B]:=AB-(-1)^{deg(A)deg(B)}BA$, and we recall that $(A \otimes B)(C \otimes D) = {(-1)}^{deg(B)deg(C)}  AC \otimes BD$}:
\beq
\Delta W = \Delta([ J,D]) = [\Delta (J),\Delta( D)].
\eeq
 This equality holds iff 
\beq
B(D) =  B(W).
\eeq
Similarly, if $D_1,D_2\in \mathfrak{su}(2|2)/ \mathfrak{su}(1|2)$, we again demand the equality 
\beq
\Delta([D_1,D_2]) = [\Delta D_1,\Delta D_2]. 
\eeq
Thus we conclude that 
\beq
B(D_1) = B(D_2)^{-1}
\eeq
whenever the commutator does not vanish. In particular, we expect the braiding to be an invertible element.
The commutation relations then imply 
\bea
B(\alg{Q}^1_2) &=& B(\alg{Q}^2_2),\\
B(\alg{R}^2_1) &=& B(\alg{S}^2_\alpha),\\
B(\alg{Q}^\alpha_2) &=& B(\alg{R}^1_2),\\
B(\alg{R}^1_2) &=& B^{-1}(\alg{S}^2_\alpha),\\
B(\alg{Q}^\alpha_2) &=& B^{-1}(\alg{S}^2_\beta).
\eea
Our result is that there can be just one braiding factor $B \equiv B(\alg{Q}^1_2)$ and its inverse, consistently with the analysis of the previous section.

Now let us derive the braiding of the central charges $\alg{P},\alg{K}$ and $\alg{C}$, which we can read off from the relations
\bea
\{\alg{Q}^\alpha_a,\alg{Q}^\beta_b\} &=& \epsilon^{\alpha\beta}\epsilon_{ab}\alg{P},\\
\{\alg{S}^a_\alpha,\alg{S}_\beta^b\} &=& \epsilon_{\alpha\beta}\epsilon^{ab}\alg{K},\\
\frac{1}{2}\left([\alg{Q}^1_1,\alg{S}^1_1]-[\alg{Q}^2_2,\alg{S}^2_2]\right)&=& \alg{C}.
\eea
We get
\bea
B(\alg{P}) &=& B(\alg{Q}^\alpha_2),\\
B(\alg{K}) &=& B(\alg{S}_\beta^2),
\eea
while $\alg{C}$ remains unbraided. This is again consistent with what we argued before.

The action of the coproduct on the braiding $B$ is determined from the co-associativity condition
\beq
(\Delta\otimes id)(\Delta (A)) = (id\otimes \Delta)(\Delta (A)),
\eeq
which has to be satisfied by every coalgebra. 
We get
\beq
\Delta(B) = B\otimes B,
\eeq
which means that our braiding is grouplike. Again, this is consistent with the physical 
requirement (\ref{nat1}),(\ref{nat2}).

\subsection{The Antipode and the Counit}

The antipode has to obey the equation
\beq	
\mu(S\otimes id)\Delta(A) = \mu(id\otimes S)\Delta(A) = i\circ\epsilon(A).
\eeq
Here $\epsilon$ denotes the counit, which is given by $\epsilon(A)=0\quad\forall A\in\mathfrak{su}(2|2)\ltimes {\mathbb{R}}^2$, $i$ is the unit and $\mu$ the multiplication. We recall that if a bialgebra has an antipode, then this is unique. 
For $J\in \mathfrak{su}(1|2)$ the antipode is the trivial one:
\beq
S(J)=-J.
\eeq 
If $D\in \mathfrak{su}(2|2)/\mathfrak{su}(1|2)$, we expect the braiding to appear in the antipode. Indeed, 
\bea
\mu(S\otimes id)\Delta(D) = S(D)B(D)+D = 0
\eea
gives
\beq
S(D) = -DB^{-1}(D).
\eeq
Furthermore, the action of $S$ on the braiding itself is  
\beq
S(B(D)) = B^{-1}(D).
\eeq
Using the defining coalgebra relation between the coproduct and the counit  
\beq
(id\otimes \epsilon)\Delta(A) = A = (\epsilon\otimes id)\Delta(A)
\eeq
we see that 
\beq
\epsilon(B) = 1.
\eeq

\subsection{Charge Conjugation}

For the representations we use the same convention as Beisert \cite{Beisert}. The represented generators $\pi(A)$ act on the 4-dimensional graded vector space spanned by $\kphi,\kchi,\kpsi{1},\kpsi{2}$, and the representations are labelled by the numbers $a,b,c,d$ , with the constraint $ad - bc = 1$. For the $\mathfrak{su}(1|2)$ subalgebra we have\footnote{It is understood that $A \equiv\pi(A)$}
\bea
\alg{Q}^{\alpha}_1 &=& a\, \kpsi{\alpha}\bphi + b\, \epsilon^{\alpha\beta}\kchi\bpsi{\beta}\\
\alg{S}^1_{\alpha} &=& c\, \epsilon_{\alpha\beta}\kpsi{\beta}\bchi + d\, \kphi\bpsi{\alpha}\\
\alg{R}^1_1 &=& \frac{1}{2}(\kphi\bphi - \kchi\bchi)\\
\alg{L}^\alpha_\beta &=& \kpsi{\alpha}\bpsi{\beta}- \frac{1}{2}\delta^\alpha_\beta\kpsi{\gamma}\bpsi{\gamma} 
\eea
For the generators in $\mathfrak{su}(2|2)/\mathfrak{su}(1|2)$ we have
\bea
\alg{Q}^{\alpha}_2 &=& a\, \kpsi{\alpha}\bchi - b\, \epsilon^{\alpha\beta}\kphi\bpsi{\beta}\\
\alg{S}^2_{\alpha} &=& -c\, \epsilon_{\alpha\beta}\kpsi{\beta}\bphi + d\, \kchi\bpsi{\alpha}\\
\alg{R}^1_2 &=& \kphi\bchi\\
\alg{R}^2_1 &=& \kchi\bphi .
\eea
Finally, for the centre we have
\bea
\alg{C} &=& \frac{1}{2}(ad + cb),\\
\alg{P} &=& ab,\\
\alg{K} &=& cd.
\eea

In this section we derive the charge conjugation $C$, which has to fulfill \cite{Janik}
\beq
\pi(S(A)) = C^{-1}\bar{\pi}(A)^{st}C.
\eeq
We will adopt the notation of \cite{Janik}, and we will show how it is possible to get all the parameters $\bar{a},\bar{b},\bar{c},\bar{d}$, which parametrise $\bar{\pi}$, directly from the knowledge of the full Hopf algebra.
We start with the representation of the charge conjugation matrix determined by Janik \cite{Janik}
\begin{equation}
C = \kchi\bphi\frac{a_1b_1}{\bar{a}\bar{b}}  + \kphi\bchi
 -\frac{b_1}{\bar{a}}\kpsi{2}\bpsi{1} + \frac{b_1}{\bar{a}}\kpsi{1}\bpsi{2}.
\end{equation}
Then the inverse is given by
\begin{equation}
C^{-1} = \kphi\bchi\frac{\bar{a}\bar{b}}{a_1b_1}  + \kchi\bphi
 -\frac{\bar{a}}{b_1}\kpsi{1}\bpsi{2} + \frac{\bar{a}}{b_1}\kpsi{2}\bpsi{1}.
\end{equation}
For $J\in \mathfrak{su}(1|2)$ the antipode is the trivial one and we get the equations 
\beq
C\pi(J)C^{-1} = -\bar{\pi}(J)^{st}, 
\eeq
whilst for $D\in \mathfrak{su}(2|2)/\mathfrak{su}(1|2)$ we have the relation
\beq
C\pi(D)C^{-1} = -B(D)\bar{\pi}(D)^{st}.
\eeq
We can therefore determine the parameters of $\bar{\pi}$ in terms of the parameters of $\pi$:
\begin{eqnarray}
C\alg{S}^1_{1}C^{-1}&=& -\bar{\pi}^{st}({\alg{S}^1_{1}}) 
\qquad\qquad \qquad\Rightarrow \qquad \bar{d} = -\frac{c_1b_1}{\bar{a}},\quad
\bar{c} = -\frac{d_1a_1}{\bar{b}} \nonumber \\
C\alg{Q}^1_{2}C^{-1} &=& -\pi(B(\alg{Q}^1_{2}))\bar{\pi}^{st}({\alg{Q}^1_{2}}) 
\qquad
\Rightarrow \qquad \pi(B(\alg{Q}^1_{2})) = -\frac{a_1b_1}{\bar{a}\bar{b}} \nonumber \\
C\alg{S}^2_{2}C^{-1} &=&-\bar{\pi}^{st}({\alg{S}^2_{2}})\pi(B(\alg{S}^2_{2}))
\qquad
\Rightarrow \qquad\pi(B(\alg{S}^2_{2}))= \frac{d_1\bar{a}}{b_1\bar{c}} 
 = \frac{c_1\bar{b}}{\bar{d}a_1}.
\eea
With the condition 
\beq
a_1b_1=\alpha(e^{-ip}-1) ,
\eeq
which stems from the requirement of having total zero momentum for physical states, 
and with $\pi(B(\alg{R}^2_{1})) = e^{ip}$, we finally obtain
\begin{equation}
e^{ip} = \frac{\alpha}{\alpha + a_1b_1} = -\frac{\bar{a}\bar{b}}{a_1b_1}
\qquad
\Rightarrow \qquad \bar{b} = -\frac{a_1b_1\alpha}{\bar{a}}\frac{1}{\alpha + a_1b_1}\, .
\end{equation}
This indeed coincides with eq.~(58) of Janik \cite{Janik}, who could determine $\bar{b}$ only after
imposing the crossing conditions (\ref{cr}).

\section{Conclusions}

In this paper we have shown how it is possible to extend the Hopf algebra structure discovered by Janik for the subsector $\alg{su} (1|2)$ to the full $\alg{su} (2|2)$ algebra, by determing 
the action of the coproduct and of the antipode on the remaining generators. 
The result is a nontrivial action via a central element of the Hopf algebra. This construction appears
to be novel from the mathematical point of view
and we obtain a different structure to the one familiar from quantum groups. 
However, this algebra most probably resides at the core of the problem 
of determing additional symmetries and constraints for the S-matrix of the AdS/CFT correspondence. 
We have verified the Hopf algebra axioms for this structure, and we have 
used it to determine the antiparticle representation directly from the algebra.

The next step would be to construct the correspondent universal R-matrix. The conditions of ``quasi-triangularity'' 
\begin{eqnarray}
\label{qtr}
&&(\Delta \otimes id)(\rmat) = \rmat_{13} \rmat_{23}, \nonumber \\
&&(id \otimes \Delta)(\rmat) = \rmat_{13} \rmat_{12} 
\end{eqnarray}
imply that $\rmat$ satisfies the Yang-Baxter equation, and, in the presence of an antipode, the crossing conditions
\begin{eqnarray}
\label{cr}
&&(S \otimes id)(\rmat) = \rmat^{- 1}, \nonumber \\
&&(id \otimes S^{- 1})(\rmat) = \rmat^{- 1} 
\end{eqnarray}
which in turn imply $(S \otimes S){\rmat} = \rmat$. The constraint (\ref{qtr})
is reminiscent of the ``bootstrap'' principle for a relativistic S-matrix \cite{bootstrap}, and is in fact its translation in Hopf-algebraic terms (see e.g.~\cite{Delius2}). The fact that the 
S-matrix of AdS/CFT satisfies Yang-Baxter and crossing relations is a hint that it might also satisfy the quasi-triangularity condition. 

Being expressed in a representation independent way  
purely in terms of the abstract algebra generators, the universal R-matrix 
may give us  significant
help in determining the overall scalar factor in a clean way directly in 
terms of the generators of the universal enveloping algebra.
One is likely to gain also a better understanding of the origin of the phase factor and of the whole Hopf algebra structure from the point 
of view of the string theory sigma-model. We plan to investigate these issues in a future publication.

\medskip

\bf Note added: \rm While we were finalizing this note, we received notice of the publication of 
similar results by G{\'o}mez and Hern{\'a}ndez in hep-th/0608029 \cite{GH}. The problem of having a non symmetric coproduct on the center, as mentioned in their paper, is avoided by imposing the physical requirements (\ref{nat1}),(\ref{nat2}) at the algebraic level. 

\section{Acknowledgments}

We would like to thank Gleb Arutyunov, Gustav Delius, Giancarlo De Pol, 
Romuald Janik, Michael Karowski, Eberhard Kirchberg and Nicolai Reshetikhin 
for very useful discussions and suggestions. DFG supported A.T. 
within the ``Schwerpunktprogramm Stringtheorie 1096''. This work was supported in part
by the Volkswagen Foundation.

\end{document}